# THE CHALLENGES OF CASE DESIGN INTEGRATION IN THE TELECOMMUNICATION APPLICATION DOMAIN

Luiz Fernando Capretz
Department of Electrical and Computer Engineering, University of Western Ontario,
London, N6G 1H1, Canada

*The magnitude of the problems facing the telecommunication software industry is presently at a point at which software engineers should become deeply involved. This paper presents a research project on advanced telecommunication technology carried out in Europe, called BOOST (Broadband Object-Oriented Service Technology). The project involved cooperative work among telecommunication companies, research centres and universities from several countries. The challenges to integrate CASE tools to support software development within the telecommunication application domain are discussed. A software process model that encourages component reusability, named the X model, is described as part of a software life cycle model for the telecommunication software industry.*

**Keywords:** *CASE environment, software reusability, telecommunication software, software process model, software product line.*

## 1. Introduction

With the advent of intelligent networks, telecommunication systems are becoming more sophisticated and consequently more complex. The computerization of telecommunication equipment has opened up enormous opportunities to offer the users new services and features. However, it also presents problems: for instance, how to design, deliver, reuse and manage large and complex software systems. This scenario calls for new tools and better CASE (Computer-Aided Software Engineering) environment to support the tasks involved in the construction of telecommunication software.

Within the scope of software development for the telecommunication application domain, the design of an environment that makes easy the introduction of new network services is of paramount importance. Such a platform should be able to support as many of the existing software engineering tools as possible in order to extend existing technology to the task of creating new services. Additionally, the introduction of new services must be fast (mainly in the multimedia and wireless communication arenas); thus there is a growing need for reusing already available service engineering functional components.

As far as the execution of the software development process is concerned, methodologies and their automated tools should support activities relevant to each software life cycle phase in a complementary fashion. As for the telecommunication application domain, one of the goals of methodologies and tools integration is to ensure that they are applicable to different process phases, and interact effectively in support of a well-defined software development process for service engineering. Following the trends on CASE environment research, several projects were carried out on service engineering methodologies and CASE environments, such as CASSIOPEIA (Galis, 1992), SCORE (Vaillant, 1992) and ROSA (Key, 1991).

In Europe, a group of companies, universities, and research centres from several countries have gathered to take up the challenge of improving software development productivity within the telecommunication industry. Together with the Commission of the European Communities (CEC), they have sponsored a programme called RACE (Research into Advanced Communications in Europe), and in particular, the BOOST (Broadband Object-Oriented Service Technology) project. BOOST is a CASE environment that supports the development of service engineering software. Among its aims was to encourage software reusability and tools integration to be used by the telecommunication software industry.

This paper is organized as follows: the next section presents important features that should be satisfied by a CASE environment for the telecommunication application domain. In the third section a CASE tool set called the BOOST environment is described. Integration issues, which emerge when a set of tools are put to work together, are discussed in section four. The X model, an alternative software process model that enforces software reuse, is presented in section five; and conclusions on the BOOST project are outlined in the sixth section.

## 2. Tools for the Telecommunication Application Domain

There is a widespread tendency for almost everyone to use some network services, both in business and at home. This is because the information one needs may not reside inside a local computer or local network. As business enterprises grow there is also a need for more complex services. Some of them are used on a regular basis, such as airline reservation systems, and others are used temporally, e.g. automatic election systems. Given that the service engineering market is very demanding, with frequent changes in its constraints, it is imperative to find easy, fast, and cost-effective ways to meet such demands. The reuse of already implemented services provided by existing networks in such a way that the quality of the new service is guaranteed, becomes a feasible solution.

The high cost and complexity of software development and maintenance and the growing need for reusable software components are some of the factors stimulating research into better software development methodologies and CASE environments (Capretz and Capretz, 1996). Of course, the aim of a methodology is to improve some of the quality, reliability, cost-effectiveness, maintainability, management, and engineering of software. Thus one requirement for CASE tools is that they support and promote a software development methodology by sustaining and enforcing the steps, rules, principles, and guidelines dictated by that methodology.

Even though a set of distinct tools is required, it is clear that the information manipulated by each tool is going to be interrelated. A set of tools isolated from each other will not be conducive to supporting the software development process. Thus the requirement is that the tools are integrated to permit the designer to interchange information freely between one tool and another, and also to allow the designer to switch easily between tools as the needs arise.

There is also a desire for CASE tools integration through expansible environments, driven by the demand for ever-faster development of software systems. Integrated CASE environments can help software engineers to deliver software systems on time. However, to meet these demands, an integrated CASE environment must be based on a flexible framework that provides a cost-effective tool integration mechanism, encourages portable tools, facilitates the exchange of development information, and adapts to future methodologies. In such an environment, software engineers can coherently mix and match the most suitable tools that support selected methodologies. They can then plug some tools into the environment and begin working with them.

Design is normally an iterative process, and the ability to easily navigate between different tools and notations is important to permit the designer to view concurrently different facets of software development. The ability for a designer to navigate around is also vital, as reusability is something that a CASE environment must promote. A designer must be able to browse through already-captured parts of previous designs to try to see whether any components from prior work can be reused.

## 3. BOOST – Broadband Object-Oriented Service Technology

As broadband systems evolve, existing infrastructure is getting a new lease on life through Integrated Service Digital Network (ISDN). AT&T has deployed wideband ISDN, which can deliver high-quality colour images simultaneously with voice and data across the United States, and in many other countries. Across the North Atlantic, some European companies have formed a consortium to finance the BOOST project to face the challenge of rapidly creating broadband services.

In order to equip the service engineering industry to meet market challenges, BOOST has aimed at providing a software environment for service engineering by taking the pragmatic approach of enhancing existing technology by evaluating and improving it in a series of usage trials. Specifically, the major objectives of BOOST are:

- to deliver a service engineering environment, based on the rapid enhancement of existing software engineering tools;
- to evaluate and demonstrate the feasibility of the environment through a series of trials;
- to satisfy the complex needs of the service engineering application domain;
- to ensure the early availability of service engineering tools for use by the application pilots and other interested RACE projects;
- to ensure the uptake of the environment by the telecommunication software industry.

Within the BOOST Consortium, the partners have realized that some of them provide software for the telecommunication industry and others need such software to solve their problems. Therefore they divided themselves into problem owners and software providers. The BOOST partners, their origin, and their role as problem owner or software provider, are presented in Table 1 and Table 2, respectively.

**Table 1: Problem owners.**

| Institution Name | Country |
|---|---|
| DeTeBerkom | Germany |
| SEL Alcatel | Germany |
| University of Aveiro | Portugal |
| CET | Portugal |
| Telefonica | Spain |

**Table 2: Software providers.**

| Institution Name | Country |
|---|---|
| MARI Computer Systems Ltd. | U.K. |
| IPSYS Software PLC | U.K. |
| University College of Wales | U.K. |
| Bull S.A. | France |
| Societe Francais de Genie Logiciel | France |
| GIE Emeraude | France |
| Intrasoft S.A. | Greece |
| National Technical University of Athens | Greece |
| Intecs Sistemi | Italy |
| CWI | Netherlands |

The structure of the project has been basically split into three main work-packages (WP):
- *Foundations WP*: performed the research and was responsible for inter- and intra-project cooperation.
- *Trials WP*: evaluated the tools separately during the creation (by problem owners) of real services running on real networks.
- *Tools WP*: developed the BOOST environment by integrating various software providers' tools.

The foundations work-package was mainly concerned with the definition of requirements (from both the problem owners' and software providers' viewpoints), the architecture to be used in the BOOST environment, and a relevant process model to create services. The first part of the trials work-package was concerned with producing detailed specifications of the trial scenarios:

1. DeTeBerkom trial involved a medical conferencing service on the BERKON network in Berlin (Germany). This service enabled general practitioners and consultants to communicate remotely using video-telephony and the simultaneous display of medical images (3D tomography) on each one's workstation.
2. SEL Alcatel trial provided multimedia services on an Alcatel network in Stuttgart (Germany). This trial used multimedia workstations developed in a previous project that supported video-telephony, joint cooperative working and joint document editing.
3. University of Aveiro built a Pay-per-View TV service on a local network in Aveiro (Portugal).
4. CET and Telefonica worked with creation of services on an intelligent network. They have developed a Distributed Functional Plane Platform, which was used to test and validate services prior to deployment.

In the early stages of this project, trials were mainly concerned with distribution of information related to the tools which software providers were bringing into the project, and with selection of tools by the problem owners (trial partners). The resulting selection was based on demonstration and presentation of plans for tools enhancements. There was a significant effort in packing and distributing the tools and offering help and training in the use of those tools. Based on discussions with the trial partners and their experience with the tools, the tools work-package started making some initial enhancements to the tools set.

A number of candidate tools to be included in the BOOST environment are shown in Table 3. The tool names are in the vertical axis. The horizontal axis represents the various software development activities covered by each tool. It is important to notice that at least one tool tackles each important aspect of the software development process, from requirements to configuration management.

**Table 3: BOOST tools set versus process model (PM) coverage.**

| TOOL/PM | Requirem. | Analysis | Design | Docum. | Program. | Test | Reuse | Mainten. | C.M. |
|---|---|---|---|---|---|---|---|---|---|
| SAF | ✓ | ✓ | | ✓ | | | | | |
| SAW | | ✓ | ✓ | ✓ | ✓ | | | | |
| SHAPE | ✓ | ✓ | ✓ | ✓ | | | | | |
| SSADM | | ✓ | ✓ | ✓ | ✓ | | | | |
| IE | ✓ | ✓ | ✓ | ✓ | ✓ | | | | |
| HOOD | | | ✓ | ✓ | ✓ | | | | |
| AdaNICE | | | ✓ | ✓ | ✓ | | | | |
| C-NICE | | | ✓ | ✓ | ✓ | | | | |
| JAZZ | | | ✓ | ✓ | | ✓ | | | |
| R.HOOD | | | | | | | ✓ | | |
| DEMON | ✓ | ✓ | | | ✓ | | | ✓ | |
| ORDIT | ✓ | ✓ | ✓ | ✓ | | | | | |
| DocAid | | | | ✓ | | | | | |
| Museion | | | | | | | ✓ | | |
| VCM | | | | | | | | | ✓ |

**4. Integration Issues**

CASE technology has made significant advances, but its potential is limited by integration difficulties. An integrated CASE environment must have a flexible architecture that can adapt to new methods and extend to other areas. However, to meet these demands, an integrated CASE environment must be based on a framework that provides a cost-effective tool integration mechanism, encourages portable tools, facilitates the exchange of development information, and adapts to future methodologies.

The information that a tool can capture should ideally be stored in a single database to be shared by other tools in the same environment. However, this kind of integration is limited to tools from the same provider, and, generally, data within the same project. It is also possible to loosely integrate a CASE environment with translators that import and export information between tools. In a few cases, it is even possible to link tools released by different software providers if they are agreed on data formats and interfaces. However, to get to a level higher than syntactic integration (lexical checks), it is necessary to encode the semantic of the methodology into the tools; this translates into considerable implementation effort. The idea that it is possible to simply put together tools from different sources and they will be fully integrated is clearly dangerous if not impossible, unless perhaps the tools are semantically integrated before being gathered together.

Three forms of integration must be borne in mind within the BOOST context:
1. *Data integration*: it is supported by a unified data model and a common database. The goal of data integration is to ensure that all software information in the BOOST environment is managed as a consistent whole, regardless of how parts of it are manipulated.
2. *Interface integration*: it is contemplated by a uniform user interface. The goal of interface integration is to improve the efficiency and effectiveness of the man-machine interaction by being as friendly as possible.
3. *Control integration*: it is assured by a monitor that manages inter-operation and communication among tools. The goal of control integration within BOOST is to allow a flexible combination of functionality from different tools, driven by the underlying methodologies the environment supports.

The EAST environment has been conceived to meet the following requirements: tight integration between a variety of tools, customization of working procedures and overall management of large projects. It supports data integration, interface integration and control integration. It provides the essential services to sustain other important aspects of the software development process, such as: process management, project management, configuration management, and code generation - among others. The EAST environment also provides some benefits such as: completeness and consistency checks, concurrent multi-user access and automatic document generation. On top of that, it offers process-to-process communication facilities and an encapsulation mechanism for environment expansibility.

As the EAST environment is open to the addition of outsider tools, this capability makes that environment a good platform for integration of methodologies to cover the whole software life cycle. The integration of any additional tool into the EAST environment requires writing a tools script (called a *capsule*), which activates the new tool and allows the communication between that tool and the environment. There is a written procedure for software engineers who wish to encapsulate their own tools. Because of such features, EAST has been chosen as an ideal platform to integrate the BOOST tool set, which allows expansibility in that it is possible to glue tools together semantically. This is an important feature, given that there are several tools from different software providers.

## 5. The X Model

Software reuse can be broadly defined as the use of engineering knowledge or assets from existing software systems to build new ones. This is a natural technique to increase software development productivity and holds the promise of shortening software delivery time. There are COM+ from Microsoft, Enterprise JavaBeans from SUN, Component-Broker from IBM, and CORBA, among other projects in that vein.

Nowadays, software product lines (Clements and Northrop, 2002; Donohoe, 2000) are also expected to have a significant impact on the software development productivity. Software product lines usually start from analysis of the common and the variable features supporting a product-line development, and then define a set of reusable elements that can be customized and combined into new products (Kang *et al.*, 2002; Jaaksi, 2002). A product line can be built around a set of reusable components by analysing the products to determine the common and variable features using a technique called domain analysis. The software engineer develops a product structure and implementation strategy around a set of reusable components that can be glued within an architecture used as a platform for several products. If done properly, this shift can help establish a sustainable modernization practice within the software industry.

The use of software product lines as a platform of larger systems is becoming increasingly commonplace. Shrinking budgets, accelerating rates of software component enhancements, and expanding systems requirements are all driving that idea. The shift from custom development to software family is occurring in both new development and maintenance activities. Product lines are application domain focused, based on a controlled process model, and concerned primarily with the reuse of higher-level software assets, such as requirements, designs, frameworks and components. A product line based on component-based software development has broad implications for how software engineers develop and evolve software systems, so this technique is here to stay.

Using parameters to account for differences in the products that are not expressible by just selecting alternative components, a component could be simply customized. In other cases a general component could be substantially modified to create a unique component for a specific product. More abstract components can be specialized to express key variability. Abstract components that have been implemented by using an object-oriented language often can be extended through inheritance to create a concrete component that meets a particular need. This customizability greatly expands the number of applications for which a component can be re-used; therefore a re-user should exploit different mechanisms to customize a component to a particular application domain. And, therefore a software system could be regarded as being comprised of two parts:

- *An information model* that represents the general aspects of a software system.
- *A behaviour model* that represents the application-specific parts of a software system.

The information model is composed of a global view of the static representation of components of the software system, and is built during a stage that can be named *generic design*. The behaviour model is concerned with the dynamic relationships between components, showing what objects are instantiated, how objects are composed, and how they interact in the specific application. This model is created during what can be termed *specific design*. BOOST allows the distinction between the general and the specific parts of a software system.

This distinction between the generic and specific aspects of a software system is an important characteristic. The idea of being able to classify parts of a design as generic (and hence potentially reusable assets) is a powerful reason for maintaining this distinction and indeed for spending more time on the general aspects of a software than might really be needed for a particular application.

Traditional software process models do not encourage reusability within their phases. Therefore a software process model that emphasizes the importance of reuse during software development was needed. The BOOST project introduced an alternative software process model, named the X model (Markopoulos, 1993) and depicted in Figure 1. This model enforces software reuse while software development is being carried out. The model focuses on a collection of assets that can be taken from reusable libraries as well the production of potentially reusable assets.

The X model links analysis, design, implementation, unit test, integration test, and system delivery into a framework which takes into account software development with reuse of existing components as well as production of assets for future reuse. This model addresses the mechanisms used when assets are taken from and stored into reusable libraries; it supports *development with reuse* through component assembly, as well as *development for reuse* through component cataloguing. It has been proposed as an ideal software life cycle model for the telecommunication software industry.

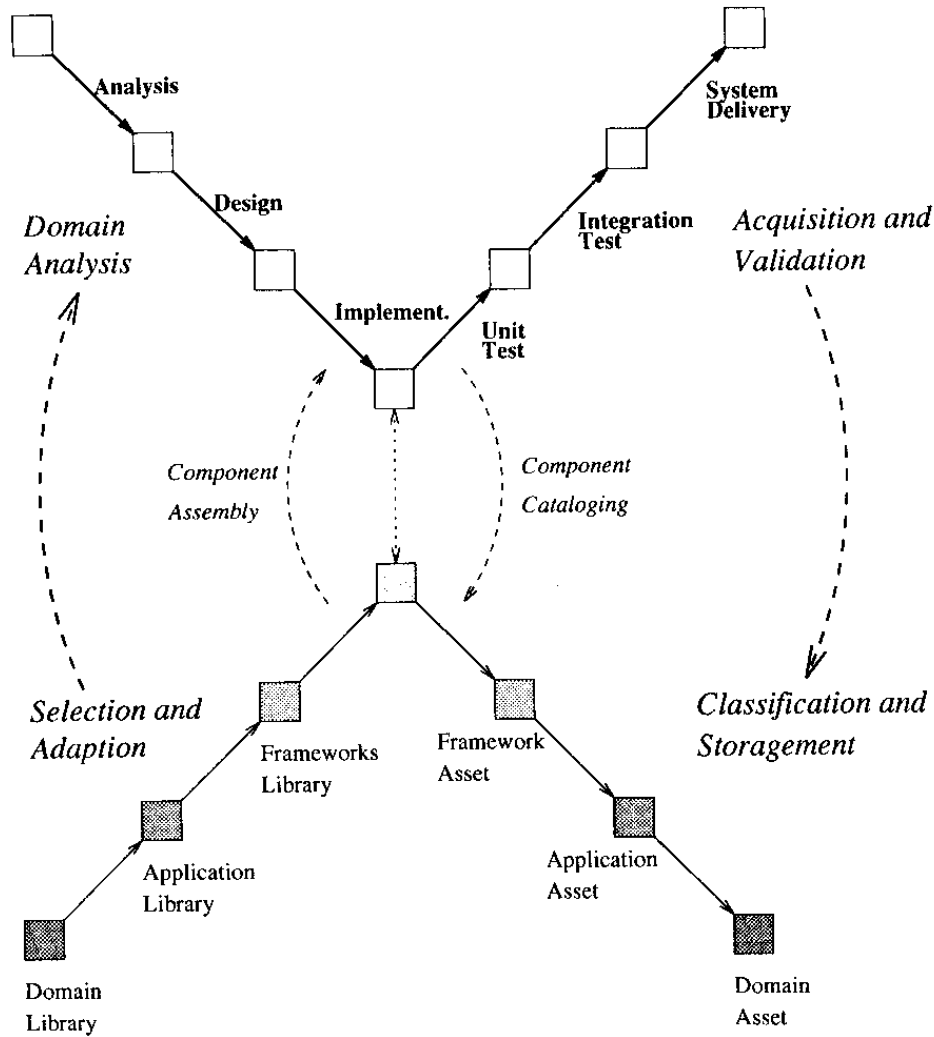

**Figure 1: The X model.**

The decisions concerning the reuse of a component are very important in that the software engineer must select the component that requires the least effort to adapt, the goal being an exact match between what is needed and what is available. Basically, the selection of a component from a reusable library involves four steps:
1. Identifying the required (target) component.
2. Selecting potentially reusable components.
3. Understanding the components.
4. Adapting (specializing, generalizing, composing or adjusting) the components to satisfy the needs of the developing software system.

The search for a component in a reusable library can lead to one of the following possible results:
- An identical match between the target and an available component is reached.
- Some closely matching components are collected, then adaptations are necessary.
- The design is changed in order to fit available components.
- No reusable component can be found; if so, the target component should be created from scratch.

While searching for components, it is necessary to address the similarity between the required (target) component and any near-matching components. The best component selected for reuse may also require specialization, generalization, or adjustment to the requirements of the new software system in which it will be reused. Sometimes, it is preferable to change the requirements in order to reuse the available components. The adaptability of the components depends on the difference between the requirements and the features offered by the existing components, as well as the skill and experience of the software designer. The process of adapting components is the least likely to become automated in the software reuse process.

One of the major problems that software engineers are faced with in trying to reuse software is the difficulty of finding reusable components once such components have been produced. This is primarily because few mechanisms are available to help identify and relate components. In order to provide more convenient reuse, the question of what kinds of mechanisms might help solve this problem arises. The answer is typically couched in terms of finding components that provide specific functionality, from a library of potentially reusable components linked through relationships that express their semantics and functionality (Capretz, 1998).

So far, most browsing tools assume that component retrieval is a simple matter of matching well-formed queries to a reusable library. But forming queries can be daunting. A software engineer's understanding of the problem evolves while searching for a component, and large reusable libraries often use an esoteric vocabulary or jargon dependent on the application domain. Therefore there is still demand for new tools to support incremental query construction to yield a flexible retrieval mechanism that satisfies ill-defined queries and reduces the terminology problem.

Tools can manipulate reusable libraries by storing, selecting and browsing the reusable components in these libraries. Selection involves browsing to find a component, retrieving it, and deploying it to the developing software system, after domain analysis is carried out. On the other hand, if a newly developed component does not exist in the reusable library, a decision has to be made as to whether the new component should be classified as a reusable component. However, before a component can be added to a reusable library, it must be validated and frozen. The validation is applied only to that particular component, not to the whole software system and should include treatment of exceptional conditions. Storing a component involves classifying it first, getting it from the developing software system, relating it to other components, and putting it into a reusable library as an asset.

## 6. Conclusions

The key issue in designing or selecting an integrated CASE environment as a platform for customized tools is how to strike the balance between integration and flexibility; tighter integration usually means less openness and expansibility. It has been taken for granted that extensible environments are a reality brought about by CASE tools and toolkit interfaces. It is true that it is possible to attach tools to particular parts of a database and share their functionality, but CASE environments only make integration easier, not simple. They are not a panacea, and they do not offer integration without a cost.

The focus of CASE research within BOOST has shifted from making sure that each tool works to ensuring that several tools can work together. Hence there can be several independent tools from different providers, but with functionality linked in such a way that these tools cover the whole software development process. The integration between such tools has been achieved through the use of a unified representation model, a uniform interface, and a monitor that establishes the communication protocol among the tools.

Finally, although many network services share a lot of functionality that can easily accommodate the requirements of a new service, software reusability has not been very common in the telecommunication industry. The BOOST project has looked at the methodologies available for service engineering creation and has attempted to find a software process model that would be suitable for telecommunication software development with reuse. The X model has been used and evaluated by major European telecommunication service providers in the context of the BOOST project. The model has proved to cope with the inherent complexity of telecommunication software. It appears to cover the likely phases of large software development and strongly supports software reuse. The experience gained in this project has been of paramount importance because component-based software design and software product lines are believed to be, in the next few years, key factors in improving software development productivity and quality.

## 7. Acknowledgements

I wish to acknowledge useful technical discussions with my colleagues at MARI Computer Systems Ltd. and other partners in this project.